\documentclass
[twocolumn,aps,prd,amsmath,amssymb,floatfix]
{revtex4-1}
\usepackage{CJK}
\usepackage[dvips]{graphicx}         %
\usepackage{bm}                      
\usepackage{mathptmx}                
\usepackage{dcolumn}                 
\usepackage{multirow}                
\usepackage{xcolor}
\usepackage[
breaklinks,pdfstartview=FitH,CJKbookmarks=true,
bookmarksnumbered=true,bookmarksopen=true,pdfborder={0 0 1},
colorlinks=true,linkcolor=blue,urlcolor=blue,anchorcolor=blue,citecolor=blue
            ]{hyperref}

\begin{document}

\def\be{\begin{equation}} \def\ee{\end{equation}}
\def\bal#1\eal{\begin{align}#1\end{align}}
\def\bse#1\ese{\begin{subequations}#1\end{subequations}}
\def\rra{\right\rangle} \def\lla{\left\langle}
\def\rv{\bm{r}} \def\tv{\bm{\tau}} \def\sv{\bm{\sigma}}
\def\tt{(\tv_1\cdot\tv_2)} \def\ss{(\sv_1\cdot\sv_2)}
\def\non{\nonumber}
\def\ra{\rightarrow}
\def\arsinh{\mathop{\text{arsinh}}}
\def\al{\alpha}
\def\la{\Lambda}
\def\eps{\epsilon}
\def\ms{\,M_\odot}
\def\mmax{M_\text{max}}
\def\l14{\Lambda_{1.4}}
\def\r14{R_{1.4}}
\def\r14{R^{(N)}_{1.4}}
\def\lr14{\l14 / \la_\text{fit}(\r14)}
\def\fm3{\;\text{fm}^{-3}}
\def\km{\;\text{km}}
\def\mev{\;\text{MeV}}
\def\gev{\;\text{GeV}}
\def\mfm{\text{MeVfm}^{-3}}
\def\fb#1{\textcolor{blue}{#1}}
\def\hcd#1{\textcolor{magenta}{#1}}
\long\def\hj#1{\color{red}#1\color{black}}
\def\OFF#1{}

\title{
Dark matter effects on the properties of quark stars and the implications for the peculiar objects
}

\begin{CJK*}{UTF8}{gbsn}

\author{Hong-Ming Liu$^1$}\email[]{liuhongming13@126.com}
\author{Peng-Cheng Chu$^1$}\email[]{kyois@126.com}
\author{He Liu$^1$} \email[]{liuhe@qut.edu.cn}
\author{Xiao-Hua Li$^{2,3}$} \email[]{lixiaohuaphysics@126.com}
\author{Zeng-Hua Li$^4$} \email[]{zhli09@fudan.edu.cn}

\affiliation{$^1$
\hbox{The Research Center for Theoretical Physics, Science School, Qingdao University of Technology,}\\
{Qingdao 266033, People's Republic of China}\\
\hbox{$^2$ School of Nuclear Science and Technology, University of South China, Hengyang,
421001, China}\\
\hbox{$^3$ Cooperative Innovation Center for Nuclear Fuel Cycle Technology \& Equipment,University of South China,}\\
{Hengyang, 421001, People's Republic of China}\\
\hbox{$^4$ Institute of Modern Physics, Key Laboratory of Nuclear Physics and Ion-Beam Application, MOE, Fudan University,} \\ 
{Shanghai 200433, People's Republic of China}
}

\date{\today}

\begin{abstract}
We systematically investigate the observable properties of dark matter-admixed quark stars (DQSs) using the confined-isospin-density-dependent-mass model in combination with the generic bosonic self-interacting dark matter model.
Our results show that the dark matter (DM) can significantly influence the properties of quark stars including the mass, radius, and the central pressure at the maximum mass configurations.
Moreover, we observe that the mass of DM particles and the DM fraction significantly affect the types of stellar configurations, and we study these configurations in detail under various scenarios and predict the possibility that two recently observed peculiar objects HESS J1731-347 and PSR J014-4002E are DQSs.

\end{abstract}

\maketitle
\end{CJK*}

\section{Introduction}

Studies on the origin and nature of dark matter (DM) are one of the most chaotic and enthralling conundrumss in particle physics and modern cosmology over the last few decades \cite{Ciarcelluti11,Henriques89,Raj18,Guever14}. This new type of particle was first proposed by Zwicky in 1933 based on galaxy cluster observations \cite{Zwicky1933} and is thought to account for approximately 27$\%$ in the Universe \cite{Begeman91,Abdalla09,Abdalla10,Wittman00,Massey10}. However, the precise characteristics of DM like its mass and interactions are presently unclear due to its very weak interactions with ordinary matter despite its existence has been supported by a plethora of evidence \cite{Henriques90,Bergstrom2000,Planck2016}. Therefore, theoretically various DM candidates have been proposed, such as the weakly interacting massive particle (WIMP) \cite{Kouvaris11, Das21,Goldman89,Andreas08,Kouvaris08,Kouvaris12,Bhat20}, asymmetric dark matter (ADM) \cite{Ivanytskyi20,Gresham19}, neutralino \cite{Hooper04,Panotopoulos17,ADas19,Das20,Das21b,Das21c,Das22,Kumar22,Lourenco22}, and axion \cite{Duffy09,Balatsky22}. From the experimental side, the observations of DM particles are mainly based on three methods, the direct detection of the cross-section between DM particles and nucleons like DAMA/LIBRA \cite{Bernabei08,Bernabei10}, CRESST-I \cite{Bravin99}, XENON \cite{XENON1T2020, Aprile18}, the indirect detection experiments through scrutinizing the products of DM candidate annihilation \cite{Abdalla10,LeDelliou15,Abdalla09} and particle accelerators produce DM \cite{Sato16,Aad2020}.            

Unfortunately, despite extensive experimental efforts, DM has not been observed until now. Compact objects such as white dwarfs, neutron stars (NSs), and strange stars could efficiently capture DM particles due to their high density and immense gravitational potential \cite{Lavallaz10,Lopes11,Bramante13,Bertoni13,Baryakhtar17,Bell21}. It is thus of great interest to unveil the features of DM by means of analyzing its effects on compact objects. NSs, as the densest objects known in the Universe, are ideal sources to probe the properties of ultra-dense matter under extreme conditions \cite{Lattimer2004,Steiner2005}. The density of NSs spans over 14 orders of magnitude from the crust to the inner core and the composition of the inner core is believed to possibly include other degrees of freedom, such as quarks \cite{PhysRevD.110.043032,PhysRevC.103.025808,Alford2019,Annala2023,PhysRevD.107.094032}, kaons \cite{PhysRevC.105.015807,PhysRevD.102.123007,Kaplan1986,PhysRevC.62.035803}, and hyperons \cite{Weissenborn2012,PhysRevC.85.065802,Chatterjee2016,Oertel2015}, alongside the basic components of neutrons, protons, electrons, and muons. According to Witten's hypothesis \cite{PhysRevD.30.272,PhysRevD.30.2379}, theoretically, NSs may transform into strange quark stars (SQSs), which are composed of absolutely stable deconfined $u$, $d$, and $s$ quark matter (QM), known as strange quark matter (SQM), along with a small number of leptons. Although there is no direct evidence for the existence of SQSs, they have not been totally ruled out as a potential explanation for the observed strange pulsars that cannot be explained by the traditional NS models. 

DM accretion affects compact objects properties in two main ways. On the one hand, DM annihilation could heat the compact object and influence its kinematic properties, such as linear and angular momentum \cite{Perezgarcia2012,PhysRevD.81.123521}. Additionally, the annihilation products, such as neutrinos or gamma rays, might be detectable by astrophysical observations and offer the indirect evidence of DM interactions. On the other hand, the capture of non-self-annihilating DM could effect the structure of compact objects. Based on the aforementioned two mechanisms, currently, many theoretical studies are now focused on exploring the effects of DM on NSs thanks to the availability of more precise and observational data \cite{Abbott17,Antoniades13,Arzoumanian18,Cromartie20,Riley21,Miller21,Pang21,Raaijmakers21,Abbott18}. Besides the fundamental effects on the NS mass-radius trajectories \cite{Leung11,Li12,Tolos15,Delpopolo20b,Das21,Yang21,Kain21}, other related phenomena, such as NSs heating due to DM annihilation or the formation and collapse of internal black holes, have been discussed and set bounds on the DM properties \cite{Kouvaris08,Kouvaris10,Gonzales10,Bertoni13,Baryakhtar17,
Garani21,Coffey22,Fujiwara22,Goldman89,Sandin09,Lavallaz10,Kouvaris12,McDermott12,
Bramante13,Bramante14,Bramante15,Ivanytskyi20,Liang23}. Moreover, more quantitative studies have been performed \cite{Das20,Biswal2021,PhysRevD.110.063001,Maselli17,Routaray23,Thakur23,Zhang22,Maselli17,Ellis18,ADas19,Nelson19,Quddus20,Husain21,Das21b,Das21c,
Das22,Leung22,Lourenco22,Karkevandi22,Dengler22,Hippert22,Collier22,Dutra22, Diedrichs23,Karkevandi23,Das21d,Emma22,Bauswein23,Zhang23,Bhat20,Kumar22,Routaray2023,PhysRevD.110.023013,PhysRevD.109.043029}. For example, the authors of Ref.~\cite{Biswal2021} investigated the impacts of DM on the curvature of the NS within the framework of relativistic mean-field theory and the results indicated the radial variation of different curvatures significantly affected by DM. In Ref.~\cite{PhysRevD.110.063001}, the authors studied the effects of DM on the nuclear saturation properties of NSs and given the constraints on the parameter space of DM particles mass and Fermi momentum by leveraging the observational data of NSs. Ref.~\cite{Das21b} proposed the possibility that the secondary component of GW190814 \cite{Abbott20b} with $M$ = 2.50$-$2.67 $\ms$ could be a DM-admixed NS.

However, we noticed that the effects of DM on the properties of SQSs has been poorly studied, which is actually very helpful in explaining some peculiar compact stars \cite{Weber2005}. SQSs are distinguished by their compactness, often exhibiting smaller radii than NSs at equivalent masses, particularly in the case of lower-mass stars \cite{Kapoor2001}. In this context, the lightest and smallest compact object so far observed by HESS Collaboration, i.~e., HESS J1731-347 \cite{Doroshenko2022} with mass $M=0.77_{-0.17}^{+0.20}$ $\ms$ and radius $R=10.4_{-0.78}^{+0.86}$ km may potentially be interpreted as a SQS because NS models struggle to explain such a small radius at such a low mass \cite{Horvath2023}. Of course, we also cannot rule out the possibility that it's a DM-admixed QS (DQS) assume that the SQSs can capure a certain amount of DM during their whole lifetime \cite{PhysRevD.109.023002}. Further multi-messenger astrophysical data will likely help clarify the nature of this compact object. Recently, Barr $et.~al$ using the MeerKAT Radio Telescope identified a companion to the pulsar PSR J0514-4002E with a mass $M$ = 2.09$-$2.71 $\ms$ at 95\% confidence interval, which falls into the so called ''mass gap" \cite{Ewan2024}.  The study suggests this could be either an exceptionally massive NS, one of the lightest black holes ever discovered or potentially an entirely new type of compact object. This paper is dedicated to explore in detail the effects of the existence of DM on the gravitational mass, radius and internal central pressure of DQSs, while probing the possibility that the companion of PSR J0514-4002E could be a DQS and determining the parameter space of DM particles mass and fraction as well as the corrsponding stellar configurations. Meanwile, we estimate DM accretion potential of another exotic compact object HESS J1731-347. For this propose, we will now assume the DM particles are the self-interacting boson and coupled to SQM only through gravitational interactions. To this end, we employ the confined-isospin-density-dependent-mass model (CIDDM) \cite{Chu_2014} and a generic bosonic self-interacting dark matter model \cite{Maselli17,Leung22,Colpi86,Chavanis12,Karkevandi22} to describe the equations of state (EOSs) of QM and DM, respectively.

This article is organized as follows.
In Sec.~\ref{Sec.2}, we briefly describe the EOSs
of ordinary QM and DM used in this work.
In Sec.~\ref{Sec.3}, The detailed calculations and discussion are presented.
A summary is given in Sec.~\ref{Sec.4}.

\section{Formalism}
\label{Sec.2}

\subsection{Equation of state for quark matter}

The EOS of dense QM of our calculations is obtain from the CIDDM model \cite{Chu_2014}, which is extended from the confined-density-dependent-mass (CDDM) model with the isospin dependence being inside the equivalent quark mass \cite{Fowler1981,Chakrabarty1989,PhysRevD.43.627,PhysRevD.48.1409,PhysRevD.51.1989,PhysRevC.61.015201}.

In the CIDDM model, the equivalent quark mass in QM with baryon density $n_B$ can be written parameterized as  
\begin{align}
m_q &= m_{q_0} + m_I + m_{\text{iso}} \notag \\
&= m_{q_0} + \frac{D}{n_B z} - \tau_q \delta D_I n^\alpha_B e^{-\beta n_B} \label{eq:my_equation}
\end{align}
where $m_{q_0}$ represents the quark current mass,  $m_I = \frac{D}{n_B z}$ accounts for the quark interactions within QM, which are presumed to density-dependent, the $z$ is a quark mass scaling
parameter, $D$ is a parameter determined based on the absolute stability condition of SQM. $m_{\text{iso}} = -\tau_q \delta D_I n^\alpha_B e^{-\beta n_b}$ is the isospin-dependent term, where $D_I$, $\alpha$ and $\beta$ are the parameters determining the isospin dependence of the quark-quark effective interactions
in QM, and $\tau_q$ is the isospin quantum number of quarks is the isospin quantum number of quarks, which are set to $\tau_q$ = 1, $-$1, and 0 for $u$, $d$, and $q$ quarks, respectively, and the isospin asymmetry $\delta$ is defined as
\bal
\delta = \frac{3 (n_d - n_u)}{n_d + n_u},
\eal
where $n_i$ is the quark number density ($i$ = $u$, $d$, and $s$ for SQM). The expression of $n_i$ can be given by
\bal
n_i = \frac{g_i}{2\pi^2} \int_0^\infty \nu_i k^2 \, dk = \frac{\nu_i^3}{\pi^2},
\eal
where $g_i$ = 6 is the value of degeneracy factor for quarks. $\nu_i$ is the Fermi momentum of different quarks and the values of $u$ and $d$ quarks can be written as 
\begin{equation}
\begin{aligned}
\nu_u &= \left(1 - {\delta}/{3}\right)^{\frac{1}{3}} \nu, \\
\nu_d &= \left(1 + {\delta}/{3}\right)^{\frac{1}{3}} \nu,
\end{aligned}
\end{equation}
where $\nu$ represents the Fermi momentum of quarks in symmetric $u$$-$$d$ QM at $n = 2n_u = 2n_d$. The total energy density of SQM then can be obtained by
\begin{equation}
\begin{aligned}
\eps_{q} = &\frac{g}{2\pi^2} \int_0^{(1-\delta/3)^{1/3} \nu} \sqrt{k^2 + m_u^2} \, k^2 \, dk \\
&+ \frac{g}{2\pi^2} \int_0^{(1+\delta/3)^{1/3} \nu} \sqrt{k^2 + m_d^2} \, k^2 \, dk \\
&+ \frac{g}{2\pi^2} \int_0^{\nu_s} \sqrt{k^2 + m_s^2} \, k^2 \, dk.
\end{aligned}
\end{equation}

In this paper we study exclusively beta-stable and neutrino-free SQM containing $u$, $d$, and $s$ quarks and electrons $e$ as relevant degrees of freedom. The composition of matter is determined by enforcing chemical equilibrium and charge neutrality at given baryon density $n_B$, which can be written as
\begin{align}
\mu_u + \mu_e = \mu_d = \mu_s,\\
\frac{2}{3} n_u = \frac{1}{3} n_d + \frac{1}{3} n_s + n_e.
\end{align}

The various chemical potentials $\mu_i$ can be obtained from the total energy density of SQM
\begin{equation}
\begin{aligned}
\mu_i =& \frac{d \eps_q}{d n_i} = \sqrt{{\nu_i^2 + m_i^2}} + \sum_j n_j \frac{\partial m_j}{\partial n_B} \frac{\partial n_B}{\partial n_i} f \left(\frac{\nu_j} {m_j}\right) \\
+&\sum_j n_j \frac{\partial m_j}{\partial \delta} \frac{\partial \delta}{\partial n_i} f \left( \frac{\nu_j} {m_j} \right)
\label{e:mu}
\end{aligned}
\end{equation}
with
\bal
f(x) = \frac{3}{2 x^3} \left[x \sqrt{\left( x^2 + 1 \right)} + \ln \left( x + \sqrt{x^2 + 1} \right)\right],
\eal
where $x={\nu_j}/{m_j}$ and the last two terms of eq.~(\ref{e:mu}) are the contributions from the density and isospin dependence of the equivalent quark mass, respectively. Especially, the chemical potential of $u$ quark $\mu_u$ can be expressed analytically as
\begin{equation}
\begin{aligned}
\mu_u = &\sqrt{\nu^2 + m_u^2} + \frac{1}{3} \sum_{j=u,d,s} n_j f \left( \frac{\nu_j} {m_j}\right)\\
& \times \left[ 
- \frac{zD} {n_B^{\left(1+z\right)}} - \tau_j D_I \delta \left( \alpha n_B^{\alpha - 1} - \beta n_B^\alpha \right) e^{-\beta n_B}\right]\\
&+ D_I n_B^\alpha e^{-\beta n_B} \left[n_u f\left(\frac{\nu_u} {m_u}\right) - n_d f\left(\frac{\nu_d} {m_d}\right)\right]\\
&\times \frac{6 n_d}{(n_u + n_d)^2}.
\end{aligned}
\end{equation}

The chemical potentials of $d$ and $s$ quarks can be written as
\begin{equation}
\begin{aligned}
\mu_d =& \sqrt{\nu^2 + m_d^2} + \frac{1}{3} \sum_{j=u,d,s} n_j f \left( \frac{\nu_j} {m_j}\right)\\
& \times \left[ 
- \frac{zD} {n_B^{\left(1+z\right)}} - \tau_j D_I \delta \left( \alpha n_B^{\alpha - 1} - \beta n_B^\alpha \right) e^{-\beta n_B}\right]\\
&+ D_I n_B^\alpha e^{-\beta n_B}\left[n_d f\left(\frac{\nu_d} {m_d}\right) - n_u f\left(\frac{\nu_u} {m_u}\right)\right] \\
&\times \frac{6 n_u}{(n_u + n_d)^2},
\end{aligned}
\end{equation}
and
\begin{equation}
\begin{aligned}
\mu_s &= \sqrt{\nu_s^2 + m_s^2} + \frac{1}{3} \sum_{j=u,d,s} n_j  f \left( \frac{\nu_j} {m_j}\right) \\
&\times  \left[ 
- \frac{zD} {n_B^{\left(1+z\right)}} - \tau_j D_I \delta \left( \alpha n_B^{\alpha - 1} - \beta n_B^\alpha \right) e^{-\beta n_B} \right].
\end{aligned}
\end{equation}

For electrons, the chemical potential is
\bal
\mu_e = \sqrt{3\pi^2 \nu_e^2 + m_e^2},
\eal
where $\nu_e$ and $m_e$ are the Fermi momentum and mass of electrons, respectively.

Therefore, the total pressure of SQM can be obtained by
\begin{equation}
\begin{aligned}
p_q =& -\eps_{q} + \sum_{j=u,d,s,e} n_j \mu_j \\
    = &- \Omega_0 +
    \sum_{i,j=u,d,s,e} n_i n_j \frac{\partial m_j}{\partial n_B} \frac{\partial n_B}{\partial n_i} f \left( \frac{\nu_j} {m_j}\right)  \\
&+ \sum_{i,j=u,d,s,e} n_i n_j \frac{\partial m_j}{\partial \delta} \frac{\partial \delta}{\partial n_i} f \left( \frac{\nu_j} {m_j}\right),
\label{e:press}
\end{aligned}
\end{equation}
 where $-\Omega_0$ is the free-particle contribution and $\Omega_0$ can be written analytically as
\[
\begin{aligned}
\Omega_0& = - \sum_{j=u,d,s,e} \frac{g_i}{48 \pi^2} \Bigg[
     \nu_i \left( \nu_i^2 + m_i^2 \right) \left( 2\nu_i^2 - 3m_i^2 \right) \\
    & + 3m_i^4 arcsinh\left(\frac{\nu_i}{m_i}\right)
\Bigg].
\end{aligned}
\]

It is worth noting that the additional terms in eq.~(\ref{e:press}) due to the density and isospin dependence of the equivalent quark mass ensure the thermodynamic self-consistency of the model and therefore fulfill the Hugenholtz$-$Van Hove theorem \cite{PhysRevC.61.015201}.

Once the QM EOS $p_q(\eps_q)$ is specified, the mass-radius relation ($M$-$R$) of a QS can be obtained by solving the well-known hydrostatic equilibrium equations of Tolman, Oppenheimer, and Volkov (TOV) \cite{PhysRev.55.374,PhysRev.55.364}. In this work, we select two parameter sets DI-85 ($D_I$ = 85 MeV $\text{fm}^{3\alpha}$ with ${\alpha}$ = 0.7 and $D$ = 22.92 MeV $\text{fm}^{-3z}$ with $z$ = 1.8) and DI-3500 ($D_I$ = 3500 MeV $\text{fm}^{3\alpha}$ with ${\alpha}$ = 0.7, $D$ = 13.81 MeV $\text{fm}^{-3z}$ with  $z$ = 1.8) within the CIDDM model \cite{Chu_2014} as the QM EOSs, which can give the maximum masses of static QSs $\mmax=2.34\ms$ and $\mmax=2.01\ms$, which all fulfill the current astronomical constraints $\mmax>2\ms$ \cite{Antoniades13,Arzoumanian18,Cromartie20}.

\subsection{Equation of state for dark matter}

In this work we assume the DM particles are massive self-interacting bosons and employ the frequently used generic DM model with only one paremeter $\epsilon_0$ to describe the DM EOS, which can be given by \cite{Maselli17,Leung22,Colpi86,Chavanis12,Karkevandi22}
\bal
p_d = \frac{4}{9}\eps_0\left(\sqrt{\frac{3}{4}\frac{\eps_d}{\eps_0} + 1} -1\right)^2,
\eal
where $p_d$ and $\eps_d$ are the pressure and energy density of pure DM, respectively. $\eps_0 = {m_d}^4/4\lambda$ with $m_d$ being the DM particles mass. Considering that the nature of DM is still unclear, there is basically no limit to the values of $m_d$ \cite{Kouvaris11,Bramante13}. We choose $m_d$ of the order of the nucleon mass, which have more obvious effects on the typical NS observations \cite{Liu23,PhysRevD.110.023024}. In the following we therefore select the four cases $m_d$ = 100, 200, 500 and 1000 MeV for further detailed study.

$\lambda$ is a dimensionless coupling constant that may be related to the DM self-interaction cross section $\sigma = \lambda^2/16\pi{m_d}^2$ \cite{Collier22,Eby16,XYLi12}. In the current work we select $\lambda = \pi$ as Ref.~\cite{Liu23}, the exact value is of less importance due to $\lambda$ entering with fourth power in the definition of the single scale parameter $\epsilon_0$. Note, that the positive $\lambda$ indicates the repulsive self interaction between DM particles, allowing for resistance to a strong enough gravitational pull and thus stabilizes pure boson dark stars.
 
In the Newtonian limit, this EOS can be expressed in the following forms:

for low density,
\bal
p_d = \frac{{\eps_d}^2}{16 \eps_0},
\eal

for high density,
\bal
p_d = \frac{1}{3}\eps_d.
\eal

\subsection{Hydrostatic configuration}
\label{Sec.2.C}

Assuming the QM and DM are coupled only through gravity, 
the stable configurations of the DQSs are obtained from a
two-fluid version of the TOV equations
\cite{Kodama72,Comer99,Sandin09}:
\bal
 \frac{dp_d}{dr} &= -[p_d + \eps_d]\frac{d\nu}{dr} \:,
\\
 \frac{dp_q}{dr} &= -[p_q + \eps_q]\frac{d\nu}{dr} \:,
\label{e:tovn}
\\
 \frac{dm}{dr}   &= 4\pi r^2 \eps \:,
\label{e:tovm}
\\
 \frac{d\nu}{dr} &= \frac{m + 4\pi r^3p}{r(r - 2m)} \:,
\label{e:tovnu}
\eal
where $r$ is the radial coordinate from the center of the star, and
$p=p_q+p_d$,
$\eps=\eps_q+\eps_d$,
$m=m_q+m_d$
are the total pressure, energy density, and enclosed mass, respectively.

The total gravitational mass of the DQS, and the DM mass fraction are 
\be
 M = m_q(R_q) + m_d(R_d) \:,
 \label{eq:gravmass}
\ee

\be
 f = \frac{m_d(R_d)}{M} \:,
 \label{eq:gravmass}
\ee

where the stellar radii $R_q$ and $R_d$
are defined by the vanishing of the respective pressures.
$R_q \neq R_d$ in the general case, there are thus two types of scenarios:
DM-core ($R_d<R_q$) or DM-halo ($R_d>R_q$) stars.

\section{Results}
\label{Sec.3}

\subsection{Equations of state}
\begin{figure}[t]
\vskip-12mm
\centerline{\hskip3mm\includegraphics[scale=0.50]{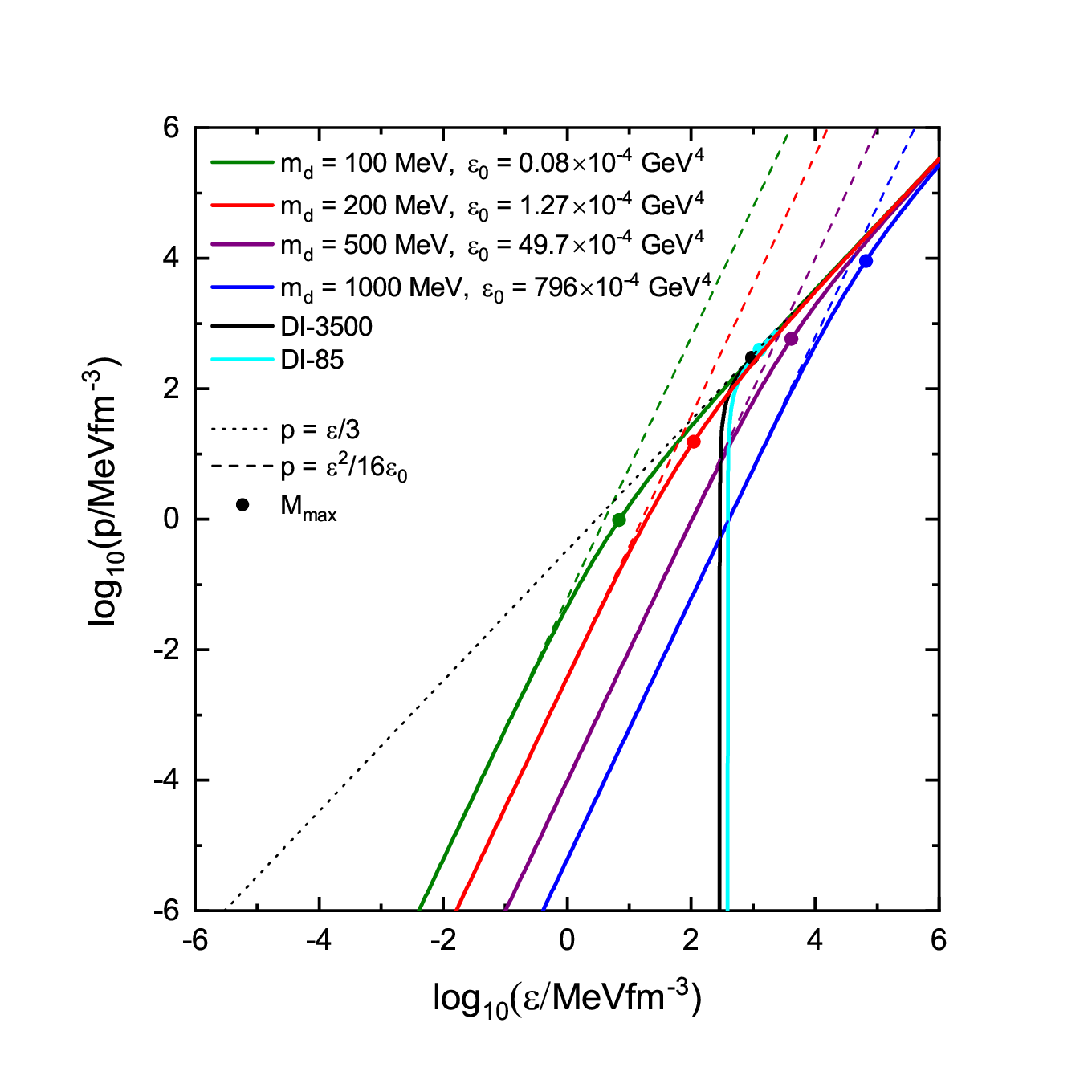}}
\vskip-9mm
\caption{
Pure DM EOSs for different DM particle masses 
and the QM DI-3500 and DI-85 EOSs.
The markers indicate the values of the maximum mass $\mmax$ configurations.
The low-density and high-density asymptotics are also
shown.
\hj{}
}
\label{f:eos}
\end{figure}

\begin{figure}[t]
\vskip-12mm
\centerline{\hskip-1mm\includegraphics[scale=0.35]{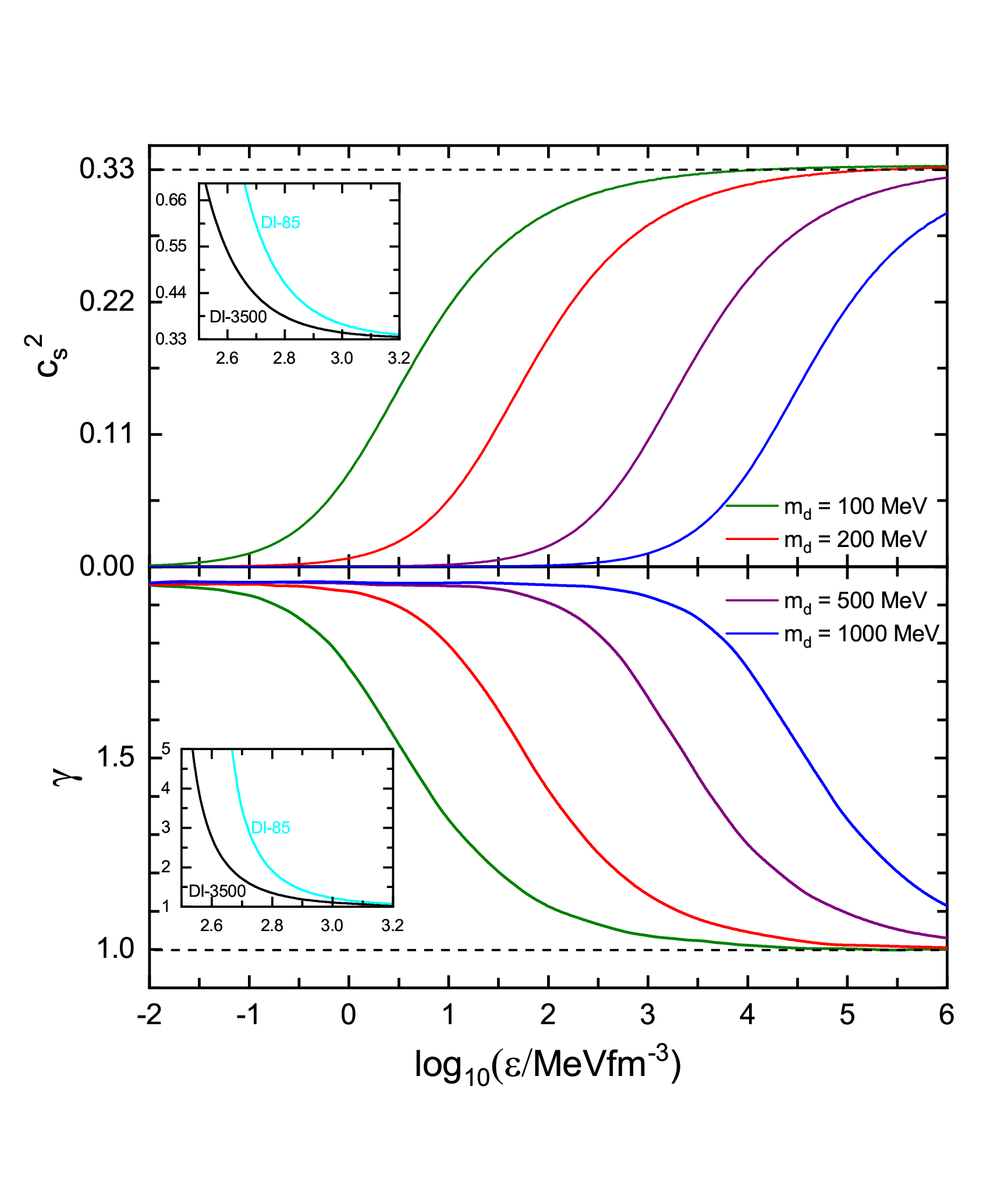}}
\vskip-10mm
\caption{
The square of the sound velocity and polytropic index as functions of logarithmic energy density in pure DM star (DM fraction $f$ = 1) for four typical DM particle masses.  
The insets show the cases of pure QS using DI-3500 and DI-85 EOSs within the CIDDM model.
}
\label{f:cs2}
\end{figure}

To begin with, we analyze the DM EOSs for DM particle masses $m_d$ = 100, 200, 500 and 1000 MeV in comparison with the QM EOSs (DI-3500 and DI-85) in Fig.~\ref{f:eos}. The markers indicate the values of the maximum mass $\mmax$ configurations. It can be seen that smaller values of $m_d$ correspond to smaller maximum energy densities and pressures, the maximum masses $\mmax$ of pure DM stars (DM fraction $f$ = 1) are 17.762, 4.441, 0.711, and 0.177 $\ms$ for $m_d$ = 100, 200, 500 and 1000 MeV, respectively. For pure QSs, the DI-3500 EOS is stiffer than DI-85 EOS and has a correspondingly larger maximum mass. The maximum masses of DQSs may exceed the ones of corrsponding pure QSs once the stiffness of DM EOSs are enough large. Moreover, one can also find that the ranges of all the DM EOSs fall within the respective low-density (dashed lines) and universal high-density (dotted line) asymptotics.

To further verify whether the QM and DM EOSs adhere to the causality condition, the upper panel of Fig.~\ref{f:cs2} shows the square of sound velocity ($c_s^2 = {\partial p}/{\partial \varepsilon}$) as a function of the logarithmic energy density of pure DM stars with $m_d$ = 100, 200, 500, and 1000 MeV. For comparison, we superimpose the results for pure QSs using DI-3500 and DI-85 EOSs within CIDDM model on the graph. The horizontal black dashed line marks the conformal matter limit $c_s^2$ = 1/3. As shown in this figure, larger values of $m_d$ correspond to smaller $c_s^2$ across the entire energy density range. The $c_s^2$ increases slowly with energy density at first, followed by a rapid rise, and eventually transitions to a slower increase. For all DM EOSs the values of $c_s^2$ approach the conformal matter limit $c_s^2$ = 1/3 as the energy density right interval is elongated, but this extends far beyond the positions of the maximum mass configurations as indicated by the markers in Fig.~\ref{f:eos}. In contrast, the $c_s^2$ decreases as energy density and ultimately approaches the conformal matter limit $c_s^2$ = 1/3 for QM using DI-3500 and DI-85 EOSs within CIDDM model. In this work, the calculated results for $c_s^2$ are presented only for the range of baryon density $n_B$ above the zero-pressure point,which all satisfy the causality limit ($c_s^2$ = 1). 

Additionly, the polytropic index $\gamma = {\partial(\rm{ln}\textit{p})}/{\partial(\rm{ln}\varepsilon)}$ is another physical quantity test if the selected DM and QM EOSs satisfy the causality condition. The lower panel of Fig.~\ref{f:cs2} depicts the relations between the polytropic index and the logarithmic energy density of pure DM stars with different $m_d$. The inset shows the pure QS cases using DI-3500 and DI-85 EOSs within CIDDM model and the horizontal black line indicates the conformal matter limit $\gamma$ = 1. It can be seen that $\gamma$ decreses with increasing energy density and gradually approaching the conformal matter limit $\gamma$ = 1, while larger $m_d$ values result in larger $\gamma$ for given energy density, consistent with the pure QS results reported in Ref.~\cite{PhysRevD.110.043032}. The values of $\gamma$ for DI-3500 EOS are always smaller than those for DI-85 EOS until the high energy density regions are nearly equal, which are slightly above the conformal matter limit $\gamma$ = 1 but well below the $\gamma$ = 1.75 in pure QM limit reported from Ref.~\cite{Annala2020}.

\begin{figure}[t]
\vskip-15mm
\centerline{\hskip-7mm\includegraphics[scale=0.52]{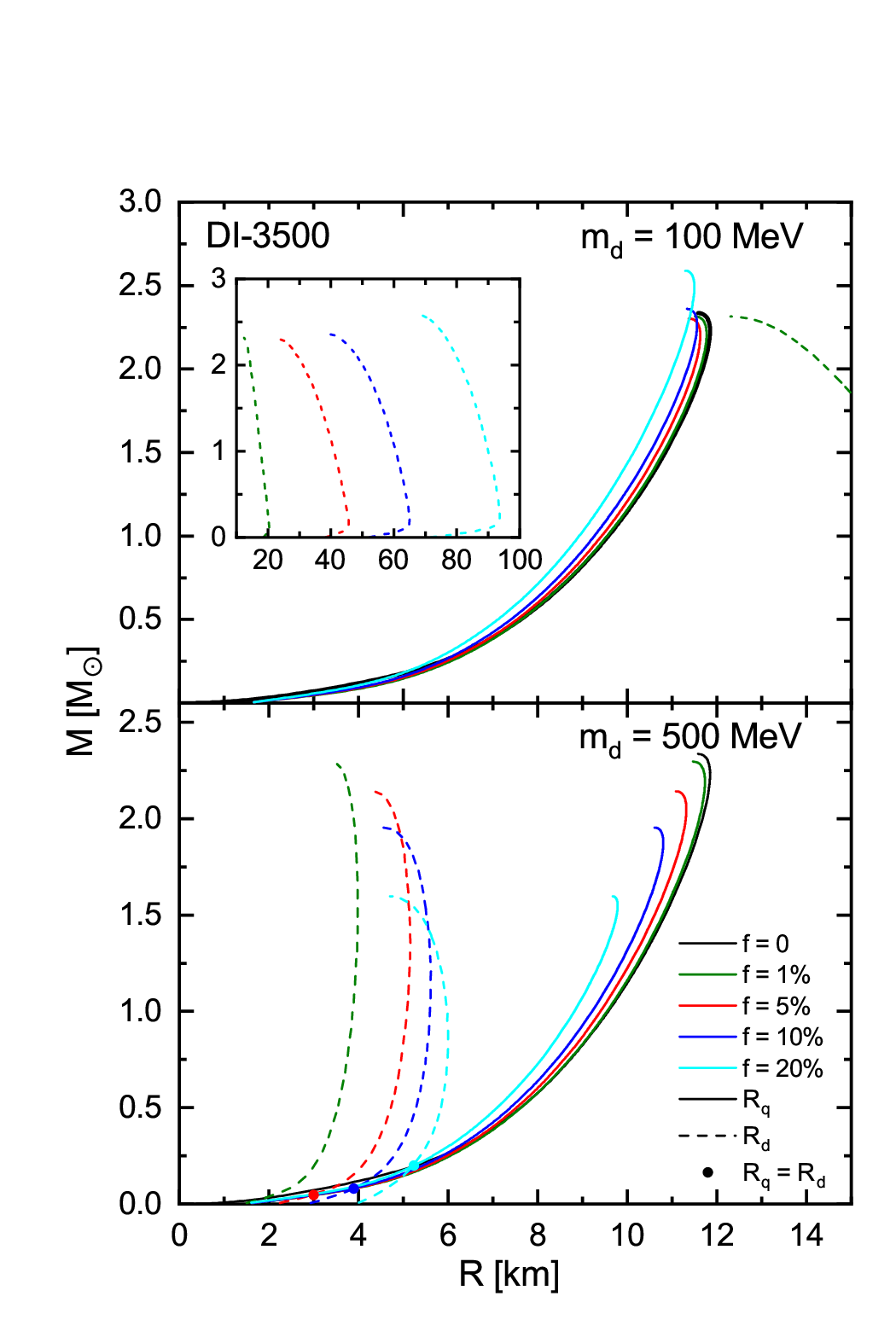}}
\vskip-6mm
\caption{
Total gravitational mass $M$ as function of QM (solid curves)
or DM (dashed curves) radius using the DI-3500 EOS
for DM particle masses $m_d$ = 100 and 500 MeV
with different DM fractions $f=M_D/M$.
Markers indicate $R_q=R_d$ configurations.
}
\label{f:dq3}
\end{figure}

\begin{figure}[t]
\vskip-14mm
\centerline{\hskip-7mm\includegraphics[scale=0.52]{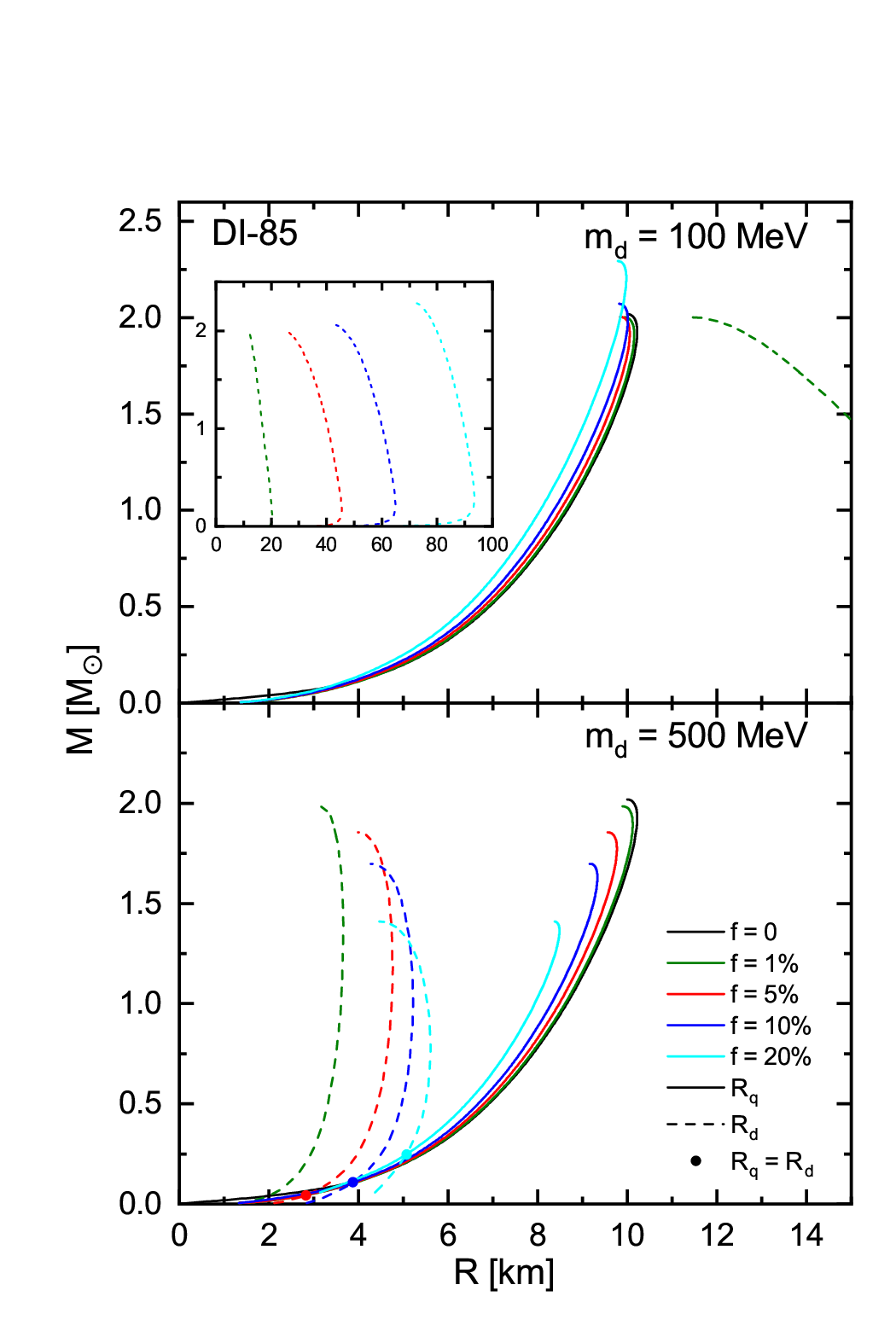}}
\vskip-6mm
\caption{Same as Fig.~\ref{f:dq3}, but for the DI-85 EOS.
}
\label{f:dq4}
\end{figure}

\begin{figure*}[t]
\vskip-12mm
\centerline{\hskip-10mm\includegraphics[scale=0.85]{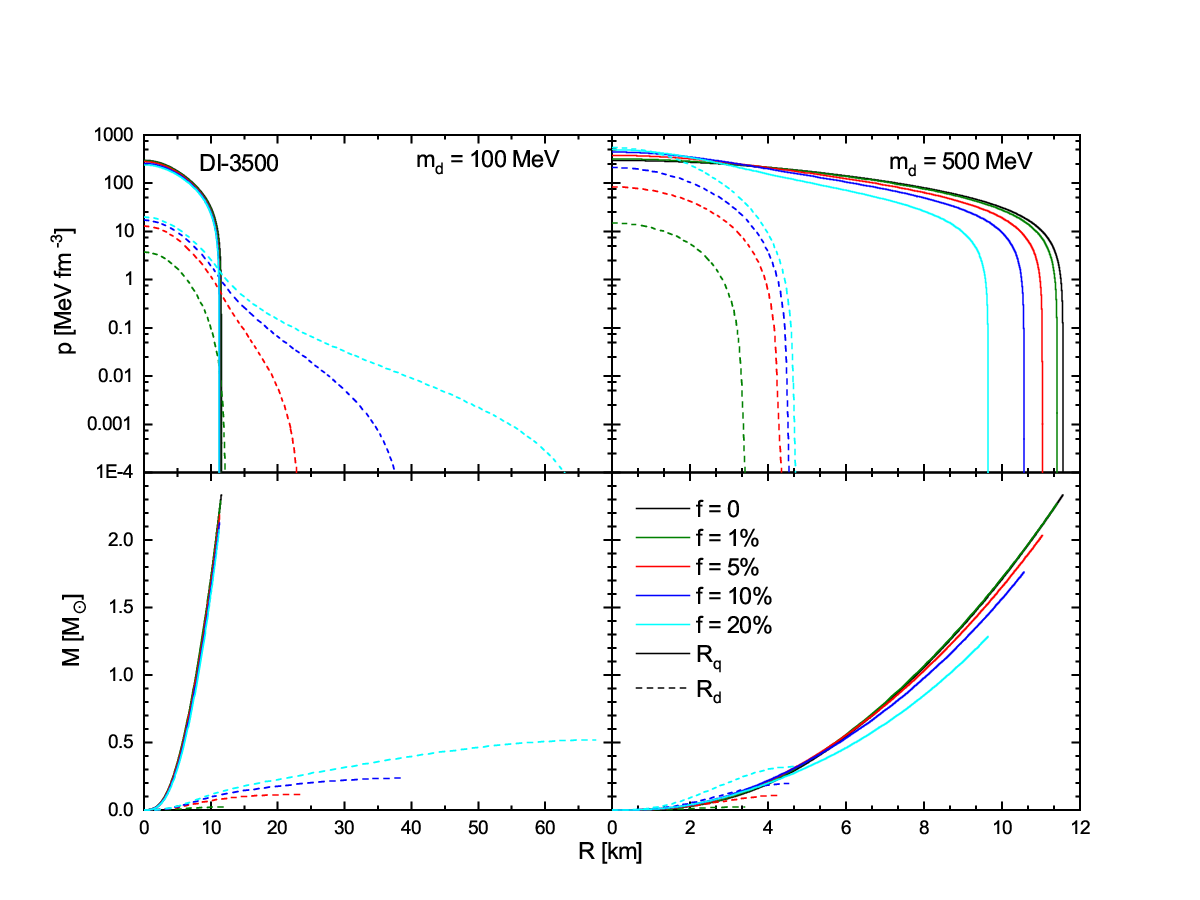}}
\vskip-6mm
\caption{
Central pressure and mass profiles of QM (solid curves) and DM (dashed curves) inside QSs at the maximum mass $\mmax$ configurations using the DI-3500 EOS
for DM particle masses $m_d$ = 100 and 500 MeV with different DM fractions $f$.}
\label{f:dq5}
\end{figure*}

\begin{figure*}[t]
\vskip-12mm
\centerline{\hskip-10mm\includegraphics[scale=0.85]{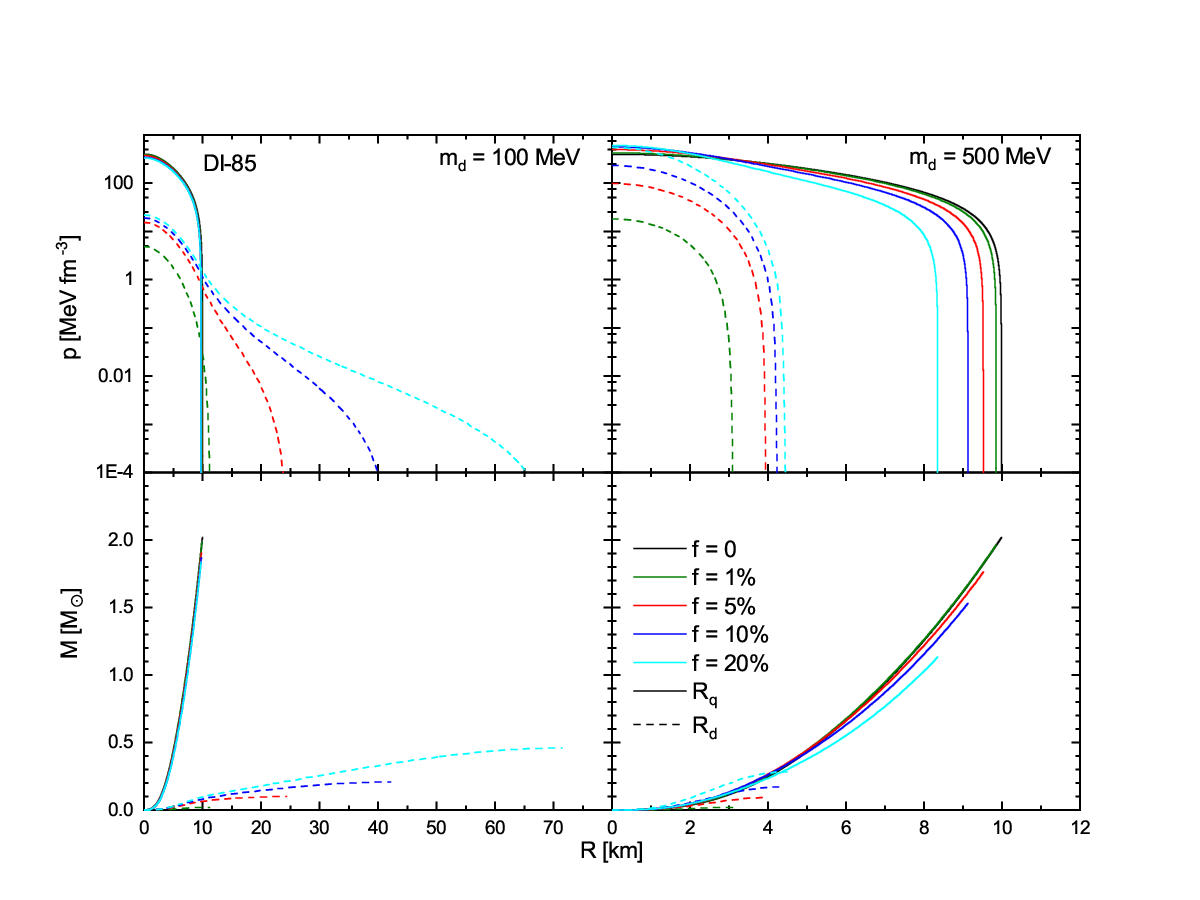}}
\vskip-6mm
\caption{
Same as Fig.~\ref{f:dq5}, but for the DI-85 EOS.
}
\label{f:dq6}
\end{figure*}

\begin{figure*}[t]
\vskip-12mm
\centerline{\hskip-1mm\includegraphics[scale=0.73]{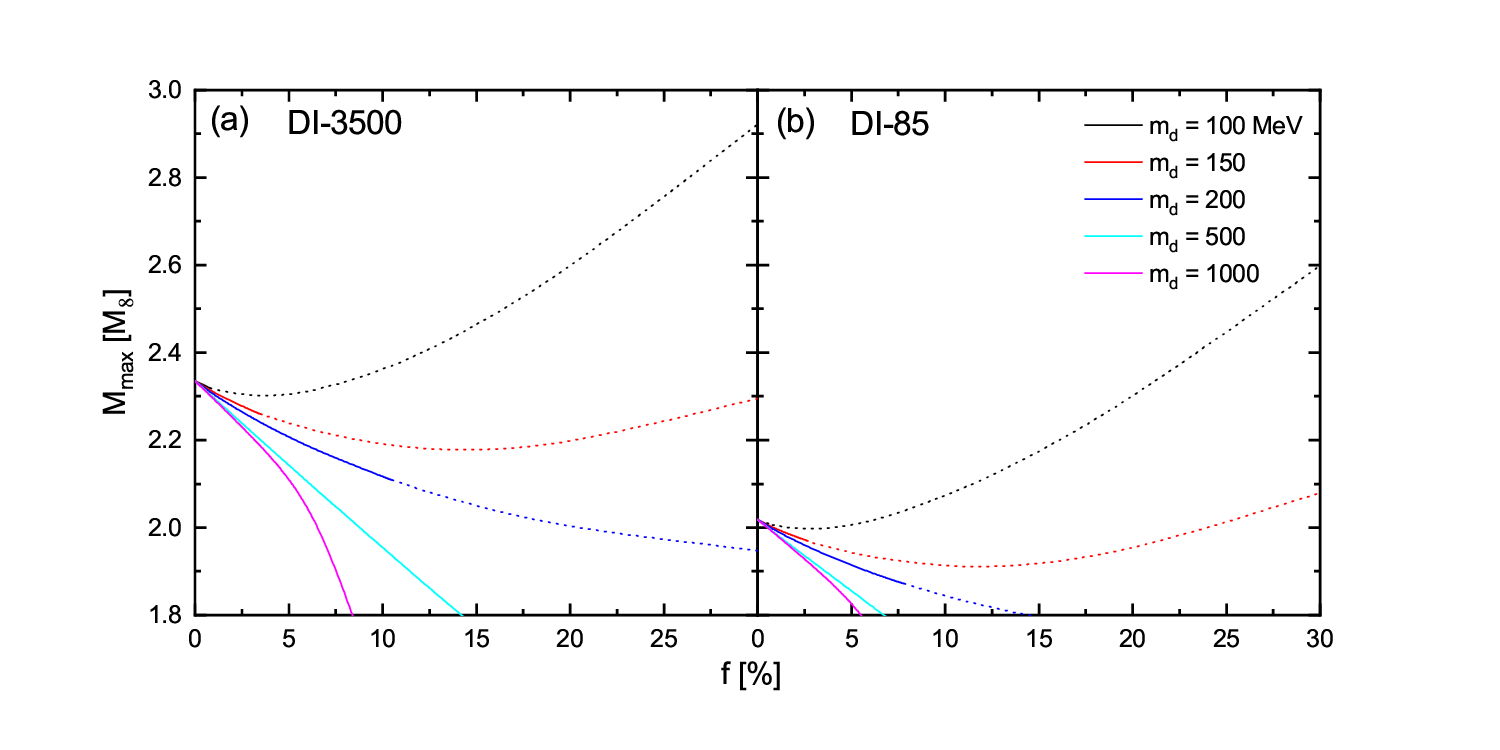}}
\vskip-10mm
\caption{
Maximum gravitational mass $\mmax$ as function of DM fractions $f$ for the DI-3500 and DI-85 EOSs with DM particle masses $m_d$ = 100, 150, 200, 500, and 1000 MeV. The DM-core and DM-halo configurations are described by the solid and dotted curves, respectively.
}
\label{f:dq7}
\end{figure*}

\begin{figure*}[t]
\vskip-8mm
\centerline{\hskip-6mm\includegraphics[scale=0.6]{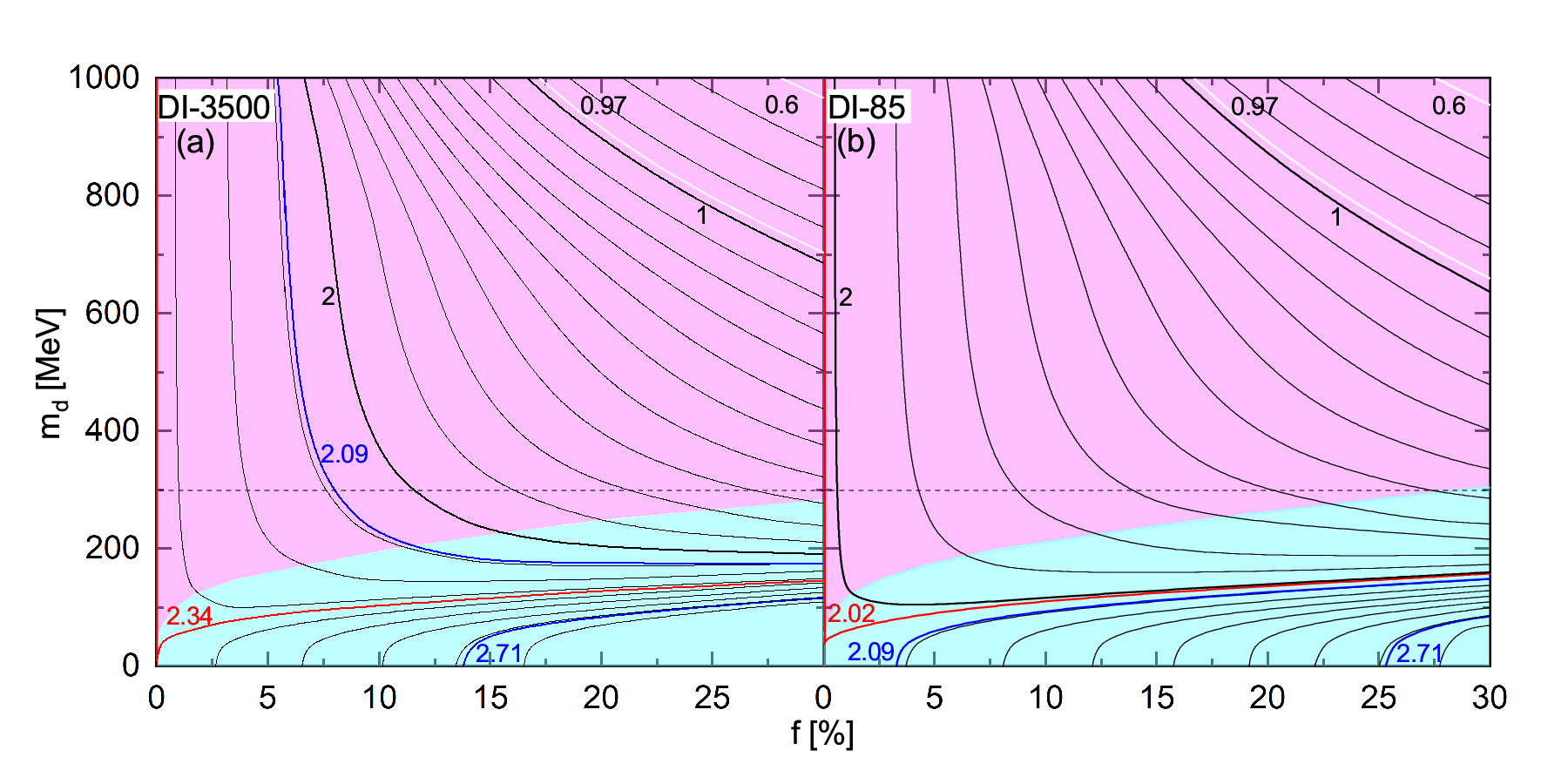}}
\vskip-4mm
\caption{Contour plots of $\mmax$ [thin contours are in intervals
of 0.1 $\ms$; 2.34 and 2.02 (red) are the values for the pure QS of DI-3500 and DI-85 EOSs; 0.6 and 0.97 (white) are the mass ranges of HESS J1731-347, 2.09 and 2.71 (blue) are the mass range of PSR J014-4002E] as functions of ($m_d$, $f$). The magenta and cyan shading domains represent the stable DM-core and DM-halo DM-admixed QSs configurations. The horizontal dashed line at $m_d$ = 300 MeV is to guide the eye.
}
\label{f:dq8}
\end{figure*}

\begin{figure*}[t]
\vskip-10mm
\centerline{\hskip-5mm\includegraphics[scale=0.75]{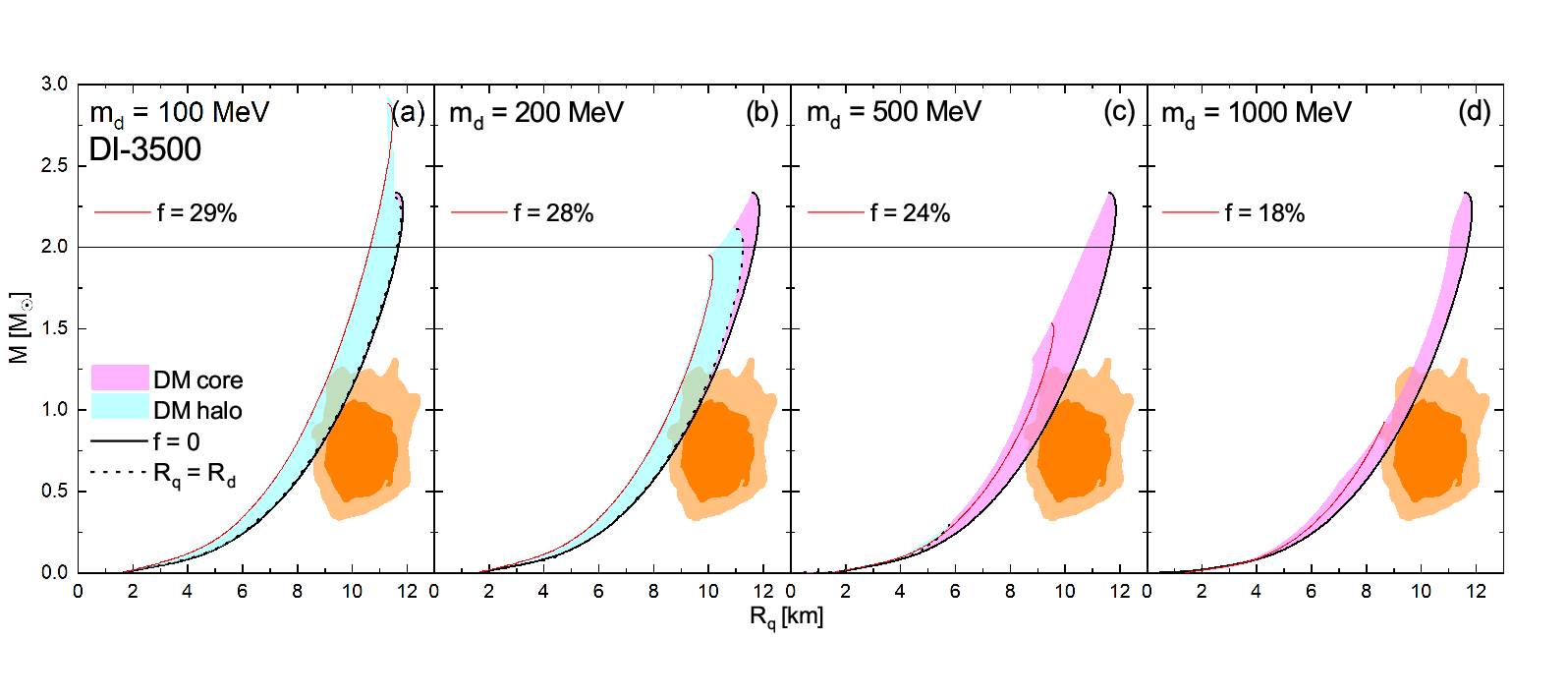}}
\vskip-10mm
\caption{
The domains of stable DM-core (magenta shading) and DM-halo (cyan shading) QSs in the $(M,R_q)$ plane
for different DM particle masses using DI-3500 EOS.
}
\label{f:dq9}
\end{figure*}

\begin{figure*}[t]
\vskip-13mm
\centerline{\hskip-5mm\includegraphics[scale=0.75]{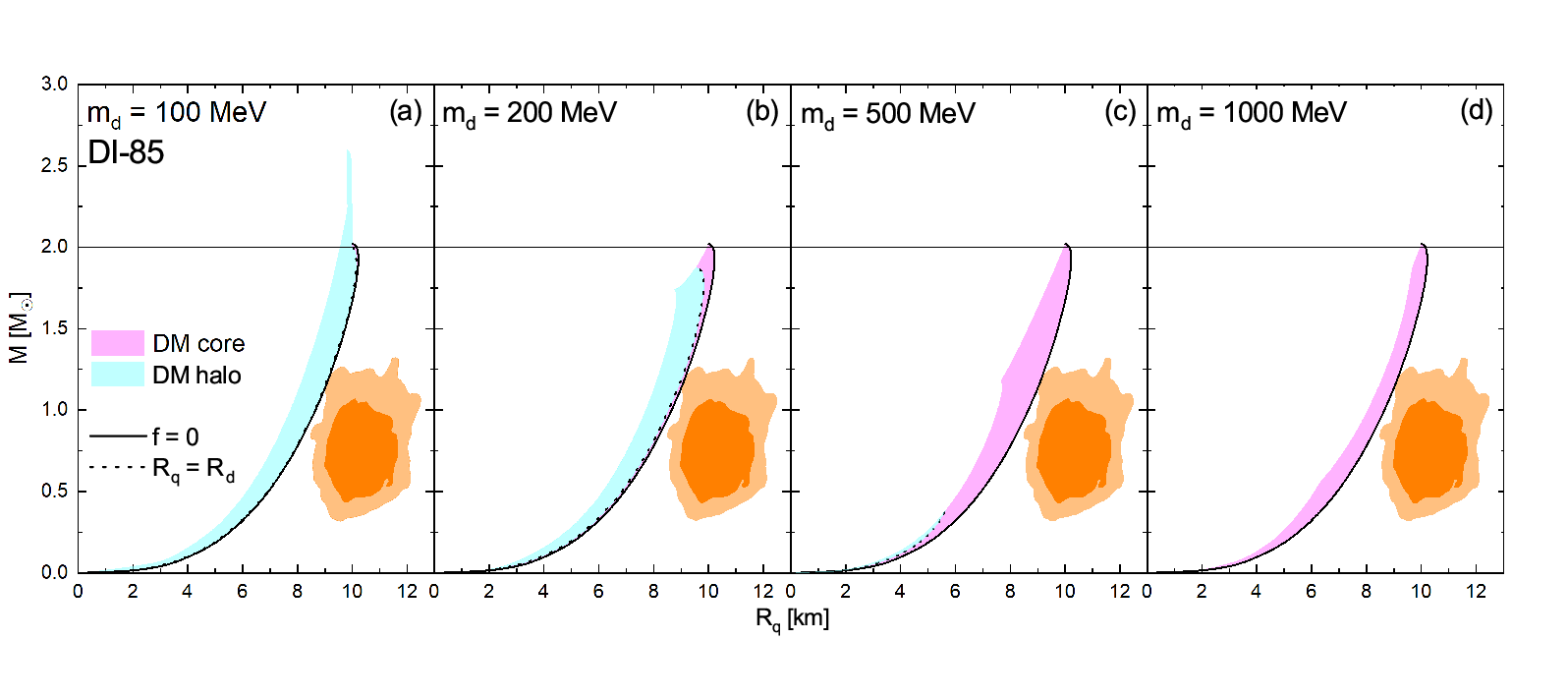}}
\vskip-12mm
\caption{
Same as Fig.~\ref{f:dq9}, but for the DI-85 EOS.
}
\label{f:dq10}
\end{figure*}

\subsection{Mass and radius}

In the following, we focus on the impact of DM on the gravitational mass and radius of DQSs in detail. Fig.~\ref{f:dq3} illustrates the total gravitational mass $M$ as functions of QM radius $R_q$ (solid curves) and DM radius $R_d$ (dashed curves) of DQSs using the DI-3500 EOS for $m_d$ = 100 (top panel) and 500 MeV (lower panel) with DM fractions $f$ = 0, 1\%, 5\%, 10\%, and 20\%. As outlined in Sec.~\ref{Sec.2.C}, DQSs can be categorized as either DM-core ($R_d<R_q$) or DM-halo ($R_d>R_q$) stars depending on the relationship between $R_q$ and $R_d$. The markers in this figure represent configurations where $R_q$ = $R_d$. It is worth noting that the accretion of DM by a star is influenced by multiple factors. For instance, whether a star resides in a high-density DM region (such as a dwarf galaxy or a DM halo) during its formation and evolution can significantly affect the total amount of DM it accretes. The density of DM is typically higher near the Galaxy center and lower in the outer regions. Consequently, the distance of a star from the Galaxy center plays a critical role in determining the amount of DM it can accumulate. In the current work, we aim to qualitatively study the effect of DM on the properties of QSs, assuming the DM fraction not exceeding 30$\%$ and without delving into the specifics of the accretion mechanism.

As we can see from Fig.~\ref{f:dq3} that the DM fraction $f$ can influence the maximum gravitational mass $\mmax$ of DQSs. For light DM particles ($m_d$ = 100 MeV), the $\mmax$ initially reduces slightly while with the further increase of $f$, $\mmax$ increases and eventually surpasses that of the pure QS ($\mmax$ = 2.34 $\ms$). However, for heavy DM particles ($m_d$ = 500 MeV), the $\mmax$ reduces consistently for any $f$. Notably, the reduction in $\mmax$ is more dramatic for heavy particles, the $\mmax$ decreases from 2.34 $\ms$ to 2.14 $\ms$ and 2.30 $\ms$ for heavy and light particles at $f$ = 5$\%$, respectively. Besides that, it can be seen that heavy DM particles can form both DM-core and DM-halo QSs, while light DM particles only form DM-halo QSs because DM radius $R_d$ is always larger than QM radius $R_q$ for any DQS mass. 

In Fig.~\ref{f:dq4}, we display the mass-radius relations of DQSs using the DI-85 EOS for $m_d$ = 100 (top panel) and 500 MeV (lower panel) at different $f$. As shown in this figure, the behavior of DI-85 EOS is qualitatively similar to that of DI-3500 EOS, but quantitatively, the decline in DI-3500 EOS is slightly larger in all cases for $m_d$ = 500 MeV. Furthermore, the increase in $\mmax$ relative to the pure QSs for DI-3500 EOS is slightly smaller than that of DI-85 EOS at $m_d$ = 100 MeV when $f$ increases from $f$ = 0 to 20\%. For example, at $f$ = 20\% (blue curves), $\mmax$ = 2.59 and 2.29 $\ms$ for DI-3500 and DI-85 EOSs with increase rates 10.7$\%$ and 13.4$\%$, respectively. 

We then turn to study the stellar interior of DQSs in detail. In Fig.~\ref{f:dq5}, we plot the central pressure (top panel) and mass (lower panel) profiles of QM (solid curves) and DM (dashed curves) of QSs at maximum gravitational mass configurations for DI-3500 EOS with different $f$. For light DM particles, the central pressure of DM $p_d$ is significantly smaller than that of QM $p_q$ and falls very slowly in contrast to $p_q$ sharply drops from its maximum to $R_q$ about 12 km. Meanwhile, the mass (radius) of DM $M_d(R_d)$ remains larger than that of QM $M_q(R_q)$ at all $f$. As an example, $R_q$ = 11.29 km and $R_d$ = 67.36 km at $f$ = 20\% (blue curves), which illustrates that DM attaches to the surface of QSs in the form of DM-halo. The increase of $f$ increases the $M_d(R_d)$ while decreases that of $M_q(R_q)$ due to the compressive effect of DM on the QM inside a QS. Once the increase in $M_d$ surpasses the decrease in $M_q$, it provides a qualitative explanation for why the total gravitational mass exceeds the pure QS maximum mass $\mmax$ = 2.34 $\ms$ when $f$ exceeds 10\% as shown in the top panel of Fig.~\ref{f:dq3}. For heavy DM particles, the $p_q$ consistently exceeds $p_d$ except at $f$ = 20\%, where the maximum values are $p_d$ = 495 and $p_q$ = 547 MeV fm$^{-3}$, respectively. The $M_q(R_q)$ is always larger than $M_d(R_d)$, indicating that DM forms a core inside a QS. With the increase of $f$, heavier DM core further attracts QM toward the interior of a QS, which raises $p_q$ and reduces $R_q$. The $\mmax$ of DQSs decreases overall because the increase in $M_d(R_d)$ is smaller than the decrease in $M_q(R_q)$. For the sake of completeness, we also display the radial profiles of the central pressure (top panel) and mass (lower panel) of QM (solid curves) and DM (dashed curves) in the maximum gravitational mass configurations of DQSs using the DI-85 EOS in Fig.~\ref{f:dq6}. The behavior of DI-85 EOS is similar to that of DI-3500 EOS across varying $f$ for both light and heavy DM particles. 

To quantify the effect of DM on the maximum mass $\mmax$ of DQSs, Fig.~\ref{f:dq7} shows the $\mmax$ of the DQSs as a function of DM fraction $f$ for (a) the DI-3500 EOS and (b) DI-85 EOS with $m_d$ = 100, 200, 500, and 1000 MeV. The solid and dotted curves show the variation of $\mmax$ with $f$ for DM-core and DM-halo configurations, respectively. It is shown that the vlaues of $m_d$ clearly influence the $\mmax$ of DQSs as well as the corresponding configurations as $f$ increases. For $m_d$ = 100 MeV, the $\mmax$ of DQSs initially decreases slightly with $f$ and then increases and exceeds that of the pure QS for both DI-3500 and DI-85 EOSs. In contrast, for $m_d$ = 150 MeV, although the behavior of $\mmax$ of DQSs with $f$ is similar to that for $m_d$ = 100 MeV, the $\mmax$ of DQSs never exceeds that of pure QSs in either the DI-3500 or DI-85 EOSs. For $m_d$ = 200, 500, and 1000 MeV, the $\mmax$ of DQSs reduces with increasing $f$ and heavier DM particles can lead to a faster reduction. Moreover, we can find that for the DI-3500 EOS the $\mmax$ of DQSs remains above the maximum mass lower limit $\mmax$ = 2 $\ms$ inferred from pulsar observations when $f \le$ 30\% for $m_d$ = 100 and 150 MeV. For $m_d$ = 200, 500, and 1000 MeV, larger values of $f$ are not allowed for fulfilling the maximum mass lower limit with the corrsponding allowed ranges being $f \le$ 20\%, 8.7\%, and 6.5\%, respectively. However, for the DI-85 EOS the allowed ranges of $f$ are narrower than those for DI-3500 for the same $m_d$ due to the smaller pure QS $\mmax$ value ($M$ = 2.02 $\ms$) of DI-85. For example, for $m_d$ = 500 MeV, the allowed $f$ can reach 8.7\% for DI-3500 EOS but only 0.5\% for DI-85 EOS when $M\ge$ 2 $\ms$. From this perspective, astronomical observations can either constrain the range of $m_d$ or $f$ if a DQS is observed in the future.

Beside that, it can be seen that the $m_d$ clearly influences the DQS configurations. The light DM particles are more likely to form DM-halo QSs, conversely heavier DM particles favor DM-core QSs. The transition from DM-core to DM-halo configurations occurs at higher $f$ as the $m_d$ increases.
\subsection{Prediction for astronomical observations}

Finally, we explore the possibility that two recent peculiar objects HESS J1731-347 with mass $M=0.77_{-0.17}^{+0.20}$ $\ms$ and radius $R=10.4_{-0.78}^{+0.86}$ km \cite{Doroshenko2022} and PSR J014-4002E with $M$ = 2.09$-$2.71 $\ms$ \cite{Ewan2024} are DQSs. To investigate this, we show firstly the contour plots of $\mmax$ of DQSs as functions of ($m_d$, $f$) plane using (a) DI-3500 EOS and (b) DI-85 EOS in Fig.~\ref{f:dq8}. The red curves represent the $\mmax$ values for pure QSs using DI-3500 and DI-85 EOSs, respectively, while the white and blue curves correspond to the mass ranges of HESS J1731-347 and PSR J014-4002E, respectively. The domains of stable DM-core (magenta shading) and DM-halo (cyan shading) DQSs are illustrated in the ($m_d$, $f$) plane. For this figure, we roughly dividing into the cases $m_d \gtrsim$ 300 MeV and $m_d \lesssim$ 300 MeV. For $m_d \gtrsim 300$ MeV, DQSs tend to form DM-core configurations by accumulating DM in the interior and the heavier DM-core can futher attract QM on the surface of QS toward the center resulting in the reductions of QM radius $R_q$ and $\mmax$. For $m_d \lesssim$ 300 MeV, the DM-halo QS configurations dominate and the $\mmax$ might reach significantly larger values with increasing $f$ (This work only presents $\mmax$ contours ranging from 0.6 to 2.8 $\ms$ and considers $f \leq 30\%$). This figure can actually be very useful for drawing any conclusion by comparison with masses of observed objects. For example, the stiff DI-3500 EOS can successfully describe the peculiar object PSR J014-4002E as either a DM-core or DM-halo QS,  as indicated by the middle region between the 2.09 $\ms$ and 2.71 $\ms$ (white) contours in the left panel of Fig.~\ref{f:dq8}. Notably, the entire considered DM particles mass range [0$-$1000 MeV] is permissible when the DM mass fraction satisfies $f \lesssim 6\%$. While with regards to the case of considering DI-85, the allowed ($m_d$, $f$) region that can explain PSR J014-4002E is much narrower than that of DI-3500 and only DM-halo QS configurations are favored, where the DM particles $m_d \gtrsim$ 148 MeV and $f \lesssim$ 3.3\% region is completely excluded. Moreover, the middle region of 0.6 $\ms$ and 0.97 $\ms$ (write) contours in the upper-right corner of the ($m_d$, $f$) plane can potentially interpret another exotic compact object HESS J1731-347 as a DM-core QS for both DI-3500 and DI-85 EOSs when $f\gtrsim$ 17\% and $m_d\gtrsim$ 700 MeV.

For futher provide the alternative explanations for the exotic compact object HESS J1731-347, we then aim to analyze how much amount of DM can be accommodated within its radius constraints of $R=_{-0.78}^{+0.86}$ km. In Fig.~\ref{f:dq9} we show the domains of stable DM-core (magenta shading) and DM-halo (cyan shading) QSs in the ($M$, $R_q$) plane for $m_d$ = 100, 200, 500, and 1000 MeV using DI-3500 EOS with $f \leq 30\%$. The shaded region in this figure represents the mass-radius confidence interval for the HESS J1731-347 and the dotted curves indicate configurations where $R_q$ =  $R_d$. Clearly, the DQSs configurations are influenced by both $m_d$ and $f$. It is worth noting that the pure QS using DI-3500 EOS ($f = 0$, black thick solid curve) can cross the shaded region, which means the HESS J1731-347 could potentially be a pure QS in the first place. As the increase of $f$, the $R_q$ reduces and $M$$-$$R_q$ curves ultimately escape the shaded region, indicating the shift from a pure QS to a DM-admixed configuration. The allowed maximum $f$ decreases with the $m_d$, which are $f$ = 29\%, 28\%, 24\%, and 18\% for $m_d$ = 100, 200, 500, and 1000 MeV, respectively. Moreover, it is noteworthy that for $m_d$ = 500 and 1000 MeV, the HESS J1731-347 could be explained by a DM-core QS within the allowable $f$ interval, while both DM-core and DM-halo QS configurations all can describe the HESS J1731-347 within the allowable $f$ interval for $m_d$ = 100 and 200 MeV.

For completeness, we also give the domains of stable DM-core (magenta shading) and DM-halo (cyan shading) QSs in the ($M$, $R_q$) plane for $m_d$ = 100, 200, 500, and 1000 MeV using the DI-85 EOS in Fig.~\ref{f:dq10}. The situation of configurations distribution for the given $m_d$ using the DI-85 EOS is similar to that of DI-3500. Both Figs.~\ref{f:dq9} and \ref{f:dq10} show that as the $m_d$ increases, the DM-core regions dominate accordingly. The pure QS using DI-85 EOS ($f = 0$, black thick solid curve) just crosses the left boundary of the shaded region, which means only a tiny amount of DM is allowed to describe the HESS J1731-347 as a DQS. In this case, any significant increase in the $f$ would push the $M$$-$$R_q$ curves outside the allowed radius region $R=_{-0.78}^{+0.86}$ km, making it less likely for HESS J1731-347 to be explained as a DQS.  In this work, we limit the parameter space to $f \leq 30\%$ and $m_d \leq 1000$ MeV. However, as the $f$ continues to increase approaches 100\% i.e., pure dark stars, the $R_q$ for a given mass may begin to increase once more, causing the $M$$-$$R_q$ curves to shift to the right into the shaded region, thereby satisfying the radius constraints of HESS J1731-347. Similar conclusions have already been extensively discussed in several works exploring DM-admixed NSs with significantly high $f$ \cite{Kain21,Liu23,PhysRevD.110.023024,PhysRevD.109.123037,PhysRevD.110.023013}. Of course, such high $f$ scenarios would necessitate exotic capture or formation mechanisms \cite{Ciarcelluti11, Goldman13, Kouvaris15, Eby16, Maselli17, Ellis18, Nelson19, Deliyergiyev19, DiGiovanni20}.

\section{Summary}
\label{Sec.4}
We have analyzed the properties of stable DQSs in a theoretical approach combining two sets of QM EOSs (stiff DI-3500 and soft DI-85) within the CIDDM model with a bosonic DM EOS involving only one parameter. Our study found that the smaller the mass of DM particles, the stiffer the DM EOS, which results in a larger pure dark star mass and a smaller polytropic index. Both the DM particles mass and DM fraction could effect the mass, radius and corrsponding configurations of DQSs. With the increase of DM fraction, the heavier DM particles tend to concentrate in the core and form DM-core QS configurations and reduce the maximum gravitational mass $\mmax$. On the contrary, for the light DM particles, the DM could form an extended halo around the QM core. In this case, DM-halo QS configurations dominate and the  $\mmax$ of DQSs initially decreases and then increases with DM fraction. The trend in the $\mmax$ of DQSs is fundamentally governed by the mass contributions and interactions between QM and DM components. This is evident from the analysis of central pressure and mass profiles of DQSs at maximum gravitational mass configurations.

In the present work we also proposed potential explanations for two recently observed peculiar compact objects: PSR J014-4002E and HESS J1731-347. By exploring the parameter space of DM particles mass and DM fraction, these objects could plausibly be interpreted as DQSs. We found that the stiff DI-3500 EOS allows for a broader parameter space for interpreting PSR J014-4002E as either a DM-core or DM-halo QS. In contrast, the soft DI-85 EOS restricts the interpretation of PSR J014-4002E to only DM-halo QS configurations with a much narrower range of allowed parameters. Regarding the compact object HESS J1731-347, both the stiff DI-3500 and soft DI-85 EOSs are capable of describing it as a DM-core QS based on its mass range. Furthermore, we analyzed how much amount of DM can be accommodated within the radius constraints of HESS J1731-347. The DI-3500 EOS can accommodate a larger amount of DM for explaining HESS J1731-347 compared to the DI-85 EOS. Meanwhile, the interpretation of HESS J1731-347 as a DQS is highly dependent on the mass of the DM particles. The heavier DM particles impose more stringent constraints on the allowed DM mass fraction in DQSs and favor DM-core configurations, whereas lighter DM particles allow for both DM-core and DM-halo configurations.

The current study provides theoretical guidance for experimental searches for DM. Combining future astronomical observations with gravitational wave signals could help to constrain DM nature further. For instance, detecting a DQS with a known DM fraction would place a lower limit on the mass of DM particles. Moreover, this study qualitatively addressed the effects of DM on QS observables and neglected the specific mechanisms for DM accretion. Future work will focus on exploring the impact of DM on the internal properties of QSs, such as the symmetry energy of QM by developing new theoretical models.

\vspace{10mm}
\section*{Acknowledgments}

This work is sponsored by the National Key R\&D Program of China No.~2022YFA1602303, the National Natural Science Foundation of China under Grant Nos.~11975132 and 12205158 and the Shandong Provincial Natural Science Foundation, China ZR2022JQ04 and ZR2021QA037.


\def\aap{A\&A}
\def\apjl{Astrophys. J. Lett.}
\def\araa{Annu. Rev. Astron. Astrophys.}
\def\epja{EPJA}
\def\epjc{EPJC}
\def\jcap{Journal of Cosmology and Astroparticle Physics}
\def\jcap{JCAP}
\def\mnras{MNRAS}
\def\npb{Nucl. Phys. B}
\def\physrep{Phys. Rep.}
\def\plb{Phys. Lett. B}
\def\ppnp{Prog. Part. Nucl. Phys.}
\def\rpp{Rep. Prog. Phys.}
\bibliographystyle{apsrev4-1}
\bibliography{dmf}

\end{document}